\documentclass[apjl]{emulateapj}

\usepackage{graphics}
\usepackage{natbib}
\usepackage{color}

\shorttitle{A Gap and a ring in the TW Hya disk}
\shortauthors{Nomura et al.}

\begin{document}


\title{ALMA Observations of a Gap and a Ring in the Protoplanetary Disk around TW Hya}


\author{Hideko Nomura\altaffilmark{1,*},
Takashi Tsukagoshi\altaffilmark{2,*},
Ryohei Kawabe\altaffilmark{3,4,5}, Daiki Ishimoto\altaffilmark{6,1}, 
Satoshi Okuzumi\altaffilmark{1}, Takayuki Muto\altaffilmark{7}, 
Kazuhiro D. Kanagawa\altaffilmark{8}, Shigeru Ida\altaffilmark{9}, 
Catherine Walsh\altaffilmark{10},
T.J. Millar\altaffilmark{11} and Xue-Ning Bai\altaffilmark{12}}



\email{nomura@geo.titech.ac.jp}


\altaffiltext{1}{Department of Earth and Planetary Sciences, Tokyo
Institute of Technology, 2-12-1 Ookayama, Meguro, Tokyo, 152-8551, Japan}
\altaffiltext{2}{College of Science, Ibaraki University, Mito, Ibaraki 310-8512, Japan}
\altaffiltext{3}{National Astronomical Observatory of Japan, 2-21-1 Osawa, Mitaka, Tokyo 181-8588, Japan}
\altaffiltext{4}{SOKENDAI (The Graduate University for Advanced
Studies), 2-21-1 Osawa, Mitaka, Tokyo 181-8588, Japan}
\altaffiltext{5}{Department of Astronomy, School of Science, University
of Tokyo, Bunkyo, Tokyo 113-0033, Japan}
\altaffiltext{6}{Department of Astronomy, Graduate School of Science,
Kyoto University, Kitashirakawa-Oiwake-cho, Sakyo-ku, Kyoto 606-8502, Japan}
\altaffiltext{7}{Division of Liberal Arts, Kogakuin University, 1-24-2 Nishi-Shinjuku, Shinjuku-ku, Tokyo, 163-8677, Japan}
\altaffiltext{8}{Institute of Low Temperature Science, Hokkaido University, Sapporo 060-0819, Japan}
\altaffiltext{9}{Earth-Life Science Institute, Tokyo Institute of
Technology, 2-12-1 Ookayama, Meguro, Tokyo 152-8550, Japan}
\altaffiltext{10}{Leiden Observatory, Leiden University, P. O. Box 9513, 2300 RA Leiden, The Netherlands}
\altaffiltext{11}{Astrophysics Research Centre, School of Mathematics
and Physics, Queen's University Belfast, University Road, Belfast BT7 1NN, UK}
\altaffiltext{12}{Institute for Theory and Computation,
Harvard-Smithsonian Center for Astrophysics, 60 Garden Street, MS-51, Cambridge, MA 02138, USA}
\altaffiltext{$^\ast$}{These authors contributed equally to this work.}


\begin{abstract}
 We report the first detection of a gap and a 
 ring in 336GHz dust continuum emission from the protoplanetary disk around TW Hya, 
 using the Atacama Large Millimeter/Submillimeter Array (ALMA). The gap and
 ring are located at around 25 and 41 AU from the central star,
 respectively, and are associated with the CO snow line at $\sim$ 30AU.
 The gap has a radial width of less than 15AU and a mass deficit of more
 than 23\%, taking into account that the observations are limited to
 an angular resolution of $\sim 15$AU. 
 In addition, the $^{13}$CO and C$^{18}$O $J=3-2$ lines
 show a decrement in CO line emission throughout the disk,
 down to $\sim$ 10AU, indicating a freeze-out of
 gas-phase CO onto grain surfaces and possible subsequent surface reactions to
 form larger molecules. 
 The observed gap could
 be caused by gravitational interaction between the disk gas and a
 planet with a mass less than super-Neptune (2$M_{\rm Neptune}$), or
 could be the result of the destruction of large dust aggregates due to the sintering
 of CO ice. 
\end{abstract}





\section{Introduction}

The physical structures and chemical compositions of gas, dust, and ice in
protoplanetary disks control the formation process of planets and the
composition of their cores and atmospheres. The 
ALMA long baseline campaign has detected gaps and rings
in dust continuum emission from the circumstellar disk around a 
very young star ($\sim 0.1-1$Myrs),
HL Tau \citep{ALMA15}. The origin of this complex disk structure 
in such a young object remains under debate. 
%

In this Letter, we present ALMA observations of the relatively old
($\sim 3-10$Myrs) gas-rich disk around the young Sun-like 
star, TW Hya ($\sim 0.8M_{\odot}$), which show the presence of a gap and a
ring associated with the CO snow line.
TW Hya's proximity (54$\pm$6 pc) makes it an ideal source to study
formation environment of a planetary system 
\citep[e.g.,][]{Andrews12}.
The disk is old compared
with other gas-rich protoplanetary disks whose lifetime is
typically $\sim 3$Myrs \citep[e.g.,][]{Hernandez07}. 
Nevertheless, the disk gas mass is $> 0.05 M_{\odot}$ 
inferred through HD line observations with {\it Herschel}
\citep{Bergin13}.
%
In exoplanetary systems giant planets have been discovered at/beyond
Neptune-orbit around Sun-like stars by direct imaging observations using
Subaru/HiCIAO \citep{Thalmann09, Kuzuhara13}. 
Thus, planets are able to form even at large distances within disk
lifetime by, for example,
scattering of planetary cores 
\citep{Kikuchi14}.
Given that 
planet-disk interaction is key for planet orbital evolution and
planet population
synthesis \citep[e.g.,][]{Kley12, Ida13},
understanding gap formation in protoplanetary disks
helps to gain insight into the early evolution
of our own Solar System, as well as the observed diversity of
exoplanetary systems.

\section{Observations and Data Reduction}

TW Hya was observed with ALMA in Band 7 on 20-21 May, 2015 
with 40 antennas in Cycle 2 
{with a uv-coverage of 22-580 k$\lambda$} (PI: D. Ishimoto). 
The spectral windows were centered at 329.295GHz (SPW1), 330.552GHz
(SPW0), 340.211GHz (SPW2), and 342.846GHz (SPW3), covering C$^{18}$O
$J=3-2$, $^{13}$CO $J=3-2$, CN $N=3-2$ and CS $J=7-6$.
The channel spacing was $\delta \nu=$30.52kHz and the bandwidth was
117.188MHz except for SPW 2 in which a channel spacing of $\delta
\nu=$15.26kHz and a bandwidth of 58.594MHz were used.
The quasar J1058+0133 was observed as a bandpass calibrator while the
nearby quasar J1037-2934 was used for phase and gain calibration.
The mean flux density of J1037-2934 was 0.58Jy during the observation
period. 

The visibility data were reduced and calibrated using 
CASA, version 4.3.1 and
4.4.0. The visibility data was reduced for each SPW
separately, and the continuum visibilities were extracted by averaging the line-free channels in all
SPWs. 
{The corrected visibilities were imaged using the CLEAN
algorithm with Briggs weighting with a robust parameter of 0 
after calibration of the bandpass, gain in amplitude and phase, 
and absolute flux scaling, and then flagging for aberrant data.} 
In addition to the usual CLEAN imaging, 
we performed self-calibration of the continuum emission
to improve the sensitivity and image quality.
The obtained solution table
of the self-calibration for the continuum emission was applied to the visibilities of the lines.
The self-calibration {significantly} improved the sensitivity of the continuum image by one order of
magnitude from 2.6 to 0.23mJy.
In addition, the continuum visibility data with deprojected baselines longer 
than 200k$\lambda$ was extracted in order to enhance small scale structure
around the gap. The high spatial resolution data was analysed in a similar 
way, but imaged using the CLEAN algorithm with uniform weighting. 
Also, the ALMA archived data of 
N$_2$H$^+$ $J=4-3$ line at 372.672GHz were reanalysed in a similar way 
to compare with that obtained by our observations.
The re-analysed data is consistent with that in \cite{Qi13}.
%
%
%
%
For the N$_2$H$^+$ line data, the synthesized beam and the 1$\sigma$ RMS noise level in 0.1 km s$^{-1}$ were 0.''44$\times$0.''41 ($P.A.=12.4^{\circ}$) and 31.2mJy beam$^{-1}$.

\section{Results}

\subsection{Dust Continuum Emission}

The observed dust continuum emission maps at 336GHz are plotted for both the
full data (Fig.~1a) and high spatial frequencies only ($> 200$k$\lambda$, Fig.~1b).
The synthesized beam and RMS noise
were 0.''37$\times$0.''31 ($\sim 20$AU, $P.A.=55.7^{\circ}$) and 0.23mJy
beam$^{-1}$. 
The total and peak flux density 
were 1.41Jy and 0.15Jy beam$^{-1}$ with a SN (signal-to-noise) ratio
of 705 and 652, respectively. 
The results agree well with the previous SMA
observations \citep{Andrews12} and ALMA observations \citep{Hogerheijde15}
with lower spatial resolution and sensitivity.
The synthesized beam and RMS for 
the data at high spatial 
resolution data only were 0.''32$\times$0.''26 ($\sim 15$AU,
$P.A.=54.5^{\circ}$) and 0.49mJy beam$^{-1}$.
%

As shown by the black line in Fig.~2a, the dust continuum emission of
the full data 
has a shallow dip in the radial profile of the surface brightness 
obtained by deprojecting the observed image data 
assuming an inclination angle of $7^{\circ}$ and a position
angle of $-30^{\circ}$ 
\citep{Qi08}. 
By imaging the data at high spatial frequencies only ($> 200$k$\lambda$),
we identify a gap and a ring at around 25AU and 41AU, respectively 
(Fig.~1b and gray line in Fig.~2a). 
{The uv-coverage of the full data set (from 22 to 580 k$\lambda$)
is sufficiently high and indeed vital for this analysis.} 
{The continuum visibility profile as a function of the deprojected
baseline length is plotted in Fig.~3. 
Our result is consistent with the recent ALMA observations with higher
spatial resolution and sensitivity \citep{Zhang16}.}
Since the high spatial frequency data misses flux of spatially extended
structure, 
the total flux
is lower than that of the full data by an order of magnitude.
Artificial structure appears at $R>70$AU in the high spatial frequency
data probably due to a failure to 
subtract the side lobe pattern of the synthesized beam. Its intensity is 
less than the 2$\sigma$-noise-level and the structure is masked in Fig.~2a.
The location of the gap is similar to that of the
axisymmetric depression in polarized intensity of
near-infrared scattered light imaging of dust grains 
recently found by Subaru/HiCIAO and
Gemini/GPI \citep{Akiyama15, Rapson15}. 

\begin{figure}
\plotone{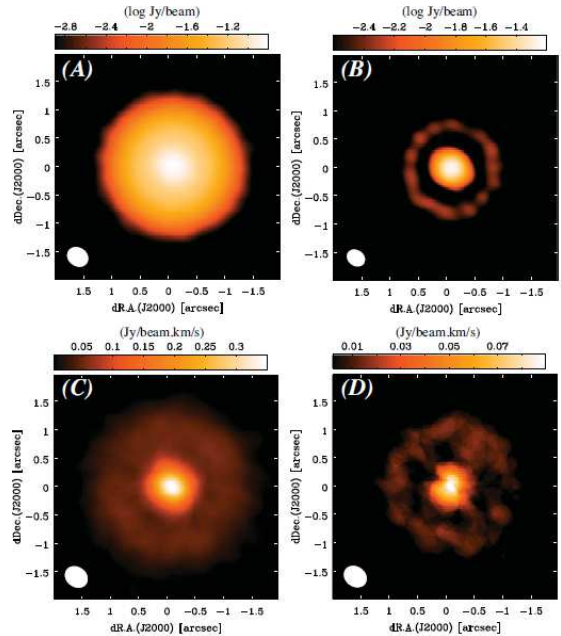}
\caption{ 
 336GHz dust continuum map imaged using (A) all spatial frequency data 
 and (B) only high spatial frequency 
 ($> 200$k$\lambda$) 
 data, (C) $^{13}$CO $J=3-2$ and
 (D) C$^{18}$O  $J=3-2$ integrated line emission maps.
Emission less than 5$\sigma$-noise-level were masked.
The synthesized beam sizes are shown in the
 bottom left corner of each panel.
\label{fig1}}
\end{figure}

\subsection{Dust Surface Density and Gap Parameters}

We extract the radial dust surface density distribution (Fig.~2b),
using the deprojected data of the dust continuum emission 
%
and the equation, 
\begin{equation}
I_{d, \nu}=B_{\nu}(T_d)[1-\exp(-\tau_{d, \nu})], \label{eq_dust}
\end{equation}
where $I_{d, \nu}$ is the observed intensity of the dust continuum
emission at the frequency $\nu$,
$B_{\nu}(T_d)$ is the Planck function 
at the dust temperature, $T_d$. 
The optical depth, $\tau_{d, \nu}$, is defined as 
$\tau_{d, \nu}=\kappa_{\nu}\Sigma_{\rm dust}$, where 
$\Sigma_{\rm dust}$ is the dust surface density and
the dust opacity, $\kappa_{\nu}$, is set as 
$\kappa_{\nu}=3.4$ cm$^2$ g$^{-1}$. 
{We adopt $T_{d}=22K(R/{\rm 10AU})^{-0.3}$, where $R$ is the disk
radius, by fitting the model result in \cite{Andrews12} and assume a
uniform temperature distribution in the vertical direction.} 
The derived dust optical depth is
shown in Fig.~2c. 
We note that the 
derived dust surface density distribution is consistent with that derived
from 
SMA observations of dust continuum
emission and model calculations \citep{Andrews12}.
Also, the assumed dust temperature is consistent with the color temperature
obtained from
the spectral index of continuum emission across the observed four
basebands between 329GHz and 343GHz (Fig.~2d) in the optically thick
regions ($\leq 30$AU, Fig.~2c).

To estimate the gap width, $\Delta_{\rm gap}$, and depth, 
$(\Sigma_0-\Sigma)/\Sigma_0$, where
$\Sigma_0$ and $\Sigma$ are the basic surface density and the
surface density with the gap at the gap center,
the intensity at the gap, $I_{\rm gap}$ (red crosses
in Fig.~2a), is derived by completing the missing flux of the
high spatial frequency data, $I^{\rm high}_{\rm gap}$ ({gray} line in
Fig.~2a), with the full
spatial frequency data, $I^{\rm full}$ (black line in Fig.~2a).
The intensity profile across the gap was obtained as follows;
(i) fit the high frequency data without the gap using a
power-law profile of $I_0^{\rm high}=1.1\times 10^3 R_{\rm AU}^{-1.5}$ mJy beam$^{-1}$,
(ii) fit the gap region in the high frequency data using a
single Gaussian profile of $\Delta I_{\rm gap}=I_0^{\rm high}-I_{\rm gap}^{\rm high}=7.7 \exp ( {-(R_{\rm AU}-24.7{\rm AU})^2}/\{2(5.9{\rm AU})^2\})$
mJy beam$^{-1}$ ({whose FWHM is $13.9\pm 1.0$AU and depth is $7.7\pm 0.5$mJy})
{(green line in Fig.~2a)}
and (iii) obtain the intensity at the gap as
$I_{\rm gap}=I^{\rm full}-\Delta I_{\rm gap}$, where $I^{\rm full}$ is
the intensity of the full data.
In this fitting, we adopted the edges of the gap at
$R=9-13$AU (inner edge) and $39-43$ AU (outer edge)
so that $\Delta I_{\rm gap}$ is overlaid to $I^{\rm full}$ smoothly. 
The optical depth at the gap is derived using $I_{\rm gap}$ and
Eq.(\ref{eq_dust}),
and then the dust surface density at the gap is derived using the same 
method mentioned above (red crosses in Fig.~2b).
The resulting gap width is $\Delta_{\rm gap} \sim 15$AU and
the depth is $(\Sigma_0-\Sigma)/\Sigma_0\sim 0.23$.
{We note that our estimate of the gap width is limited by the angular
resolution of the high 
spatial frequency data. The pattern of the gap and ring is also
affected by residual artefacts due to the cut-off at 200k$\lambda$,
which introduces uncertainties in the
location of the gap center, and the width and depth of the gap.}



\begin{figure}
\plottwo{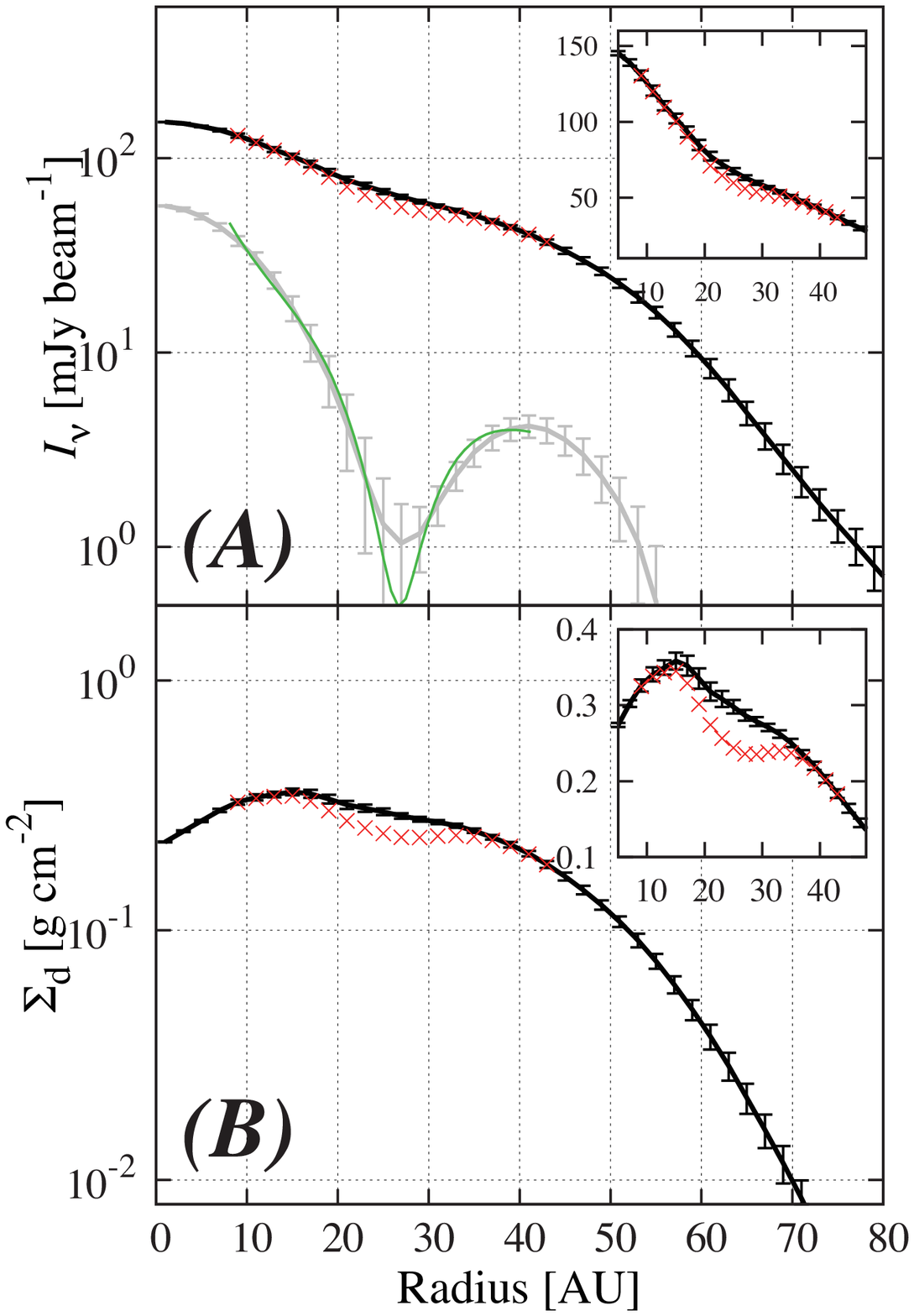}{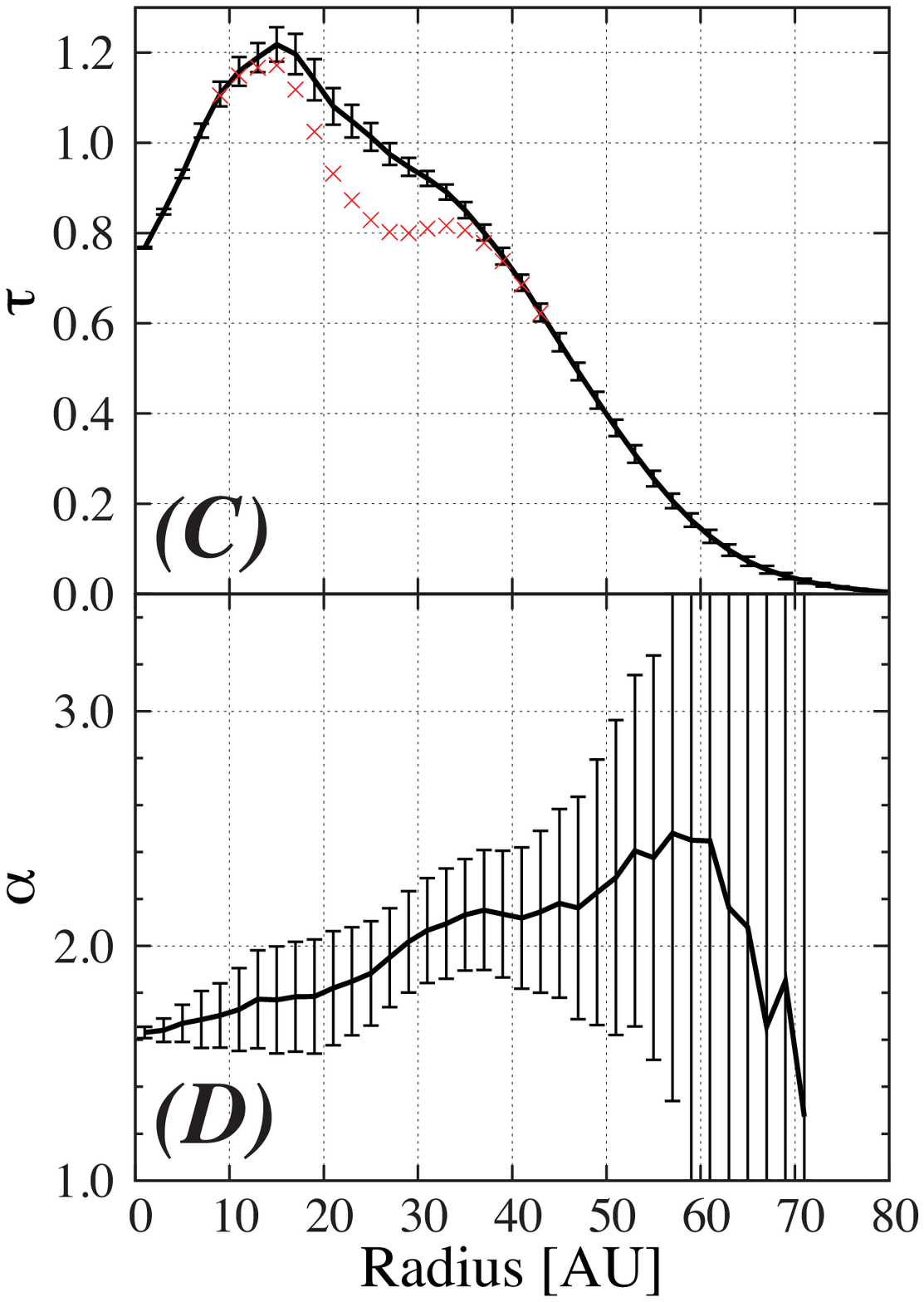}
\caption{Radial distributions of (A) intensity of 336GHz dust continuum 
 emission, (B) dust surface density, (C) optical depth and
 (D) spectral index between 329GHz and 343GHz.
 Gray line in Figure (A) shows the data of only high spatial frequency
 with its Gaussian fit (green line).
 Red crosses in (A)-(C) are derived from high spatial frequency
 data recovered by the full spatial frequency data at the gap.  
 The inserts show a close-up of the region around the gap. 
\label{fig2}}
\end{figure}

\begin{figure}[h]
\plotone{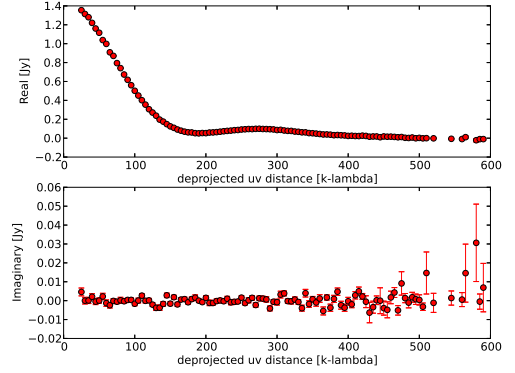}
\caption{Real (top) and imaginary (bottom) parts of the azimuthally
 averaged continuum visibility profile as a function of the deprojected
 baseline length.}
\end{figure}

\subsection{CO Isotopologue Line Emission}

The observed $^{13}$CO and C$^{18}$O $J=3-2$ rotational line emission
maps are plotted in Figs.~1c and 1d.
The resulting synthesized beam and the 1$\sigma$ RMS
noise level in 0.03 km s$^{-1}$ width-channels were
0."41$\times$0."33 ($P.A.=53.4^{\circ}$), 12.0mJy beam$^{-1}$ ($^{13}$CO)
and 0."41$\times$0."33 ($P.A.=53.8^{\circ}$), 13.4
mJy beam$^{-1}$ (C$^{18}$O). The integrated intensity maps were created by
integrating from 1.18 to 4.30 km s$^{-1}$ ($^{13}$CO)
and from 1.48 to 4.15 km s$^{-1}$ (C$^{18}$O).
The resultant noise levels of the map were 8.4 mJy beam$^{-1}$ km s$^{-1}$
($^{13}$CO) and 6.5 mJy beam$^{-1}$ km s$^{-1}$ (C$^{18}$O). 
%
%
The deprojected radial profiles of the integrated line emission of the
$^{13}$CO and C$^{18}$O {following subtraction of the dust continuum
emission} are plotted in Fig.~4a. 
The signal-to-noise ratio is not
sufficiently high to analyse the high spatial resolution data only. 

\subsection{CO Column Density}

We obtain the CO radial column density distribution, using the
deprojected data of the $^{13}$CO and C$^{18}$O line observations
(Fig.~4d). 
%
The CO column density, $N_{\rm CO}$, is derived from the integrated line
emission of the C$^{18}$O $J=3-2$ \citep{Turner91,Rybicki79},
assuming
the abundance ratios of CO $: ^{13}$CO$=$67 $:$ 1 and
$^{13}$CO $:$ C$^{18}$O$=$7 $:$ 1 \citep{Qi11}. 
%
The optical depth of the C$^{18}$O line emission, $\tau_{{\rm C^{18}O}, v}$
(Fig.~4c), is obtained from the
observed line intensity as follows.
We assume that the CO line emitting region is mainly in the surface layer, 
while the dust continuum emitting region is near the midplane. 
Thus, the deprojected intensity of the CO line emission plus dust continuum 
emission can be approximately derived from the following equation by simply 
assuming three zones in the vertical direction of the disk, 
\begin{eqnarray}
I_{{\rm C^{18}O}+{\rm cont},v} &=& B_{\nu}(T_g)(1-e^{-\tau_{{\rm C^{18}O}, v}/2}) \nonumber \\
 & & +B_{\nu}(T_d)(1-e^{-\tau_{d, \nu}})e^{-\tau_{{\rm C^{18}O}, v}/2} \\
 & & +B_{\nu}(T_g)(1-e^{-\tau_{{\rm C^{18}O}, v}/2})e^{-\tau_{d, \nu}-\tau_{{\rm C^{18}O}, v}/2}, \nonumber 
\end{eqnarray}
where $T_g$ is the gas temperature of the CO line emitting region and
$T_d$ is the dust temperature near the midplane.
The data continuum-subtracted data, $I_{{\rm C^{18}O}+{\rm cont}, v}-I_{d,\nu}$,
together with Eq.(\ref{eq_dust}) is used for the analysis. 
Since the observed line ratio of $\int I_{{\rm C^{18}O}, v}dv/\int I_{{\rm ^{13}CO},
v}dv$ is larger 
than $1/7$ (Fig.~4b), the $^{13}$CO line is optically thick.
Therefore, we adopt the brightness temperature at the peak of the 
$^{13}$CO line (Fig.~5) as the gas temperature, $T_g$.
Also, we adopt a dust temperature of $T_{d}=22K(R/{\rm 10AU})^{-0.3}$
and the dust optical depth, $\tau_{d, \nu}$, is derived from the dust 
continuum observations (Fig.~2c).
%
%
We note that in the analysis we simply assume that
{the gas temperature is uniform in the vertical direction}, and the
gas temperature of the C$^{18}$O line emitting region is the same
as that of the $^{13}$CO line emitting region, although
it could be lower in reality. 
In addition, C$^{18}$O could be depleted due to the effect of
isotopologue-dependent photodissociation, and $N_{\rm CO}/N_{\rm
C^{18}O}$ is possibly higher than 440 
\citep{Miotello14}.
The CO column density will be underestimated due to these effects.

\begin{figure}
\plottwo{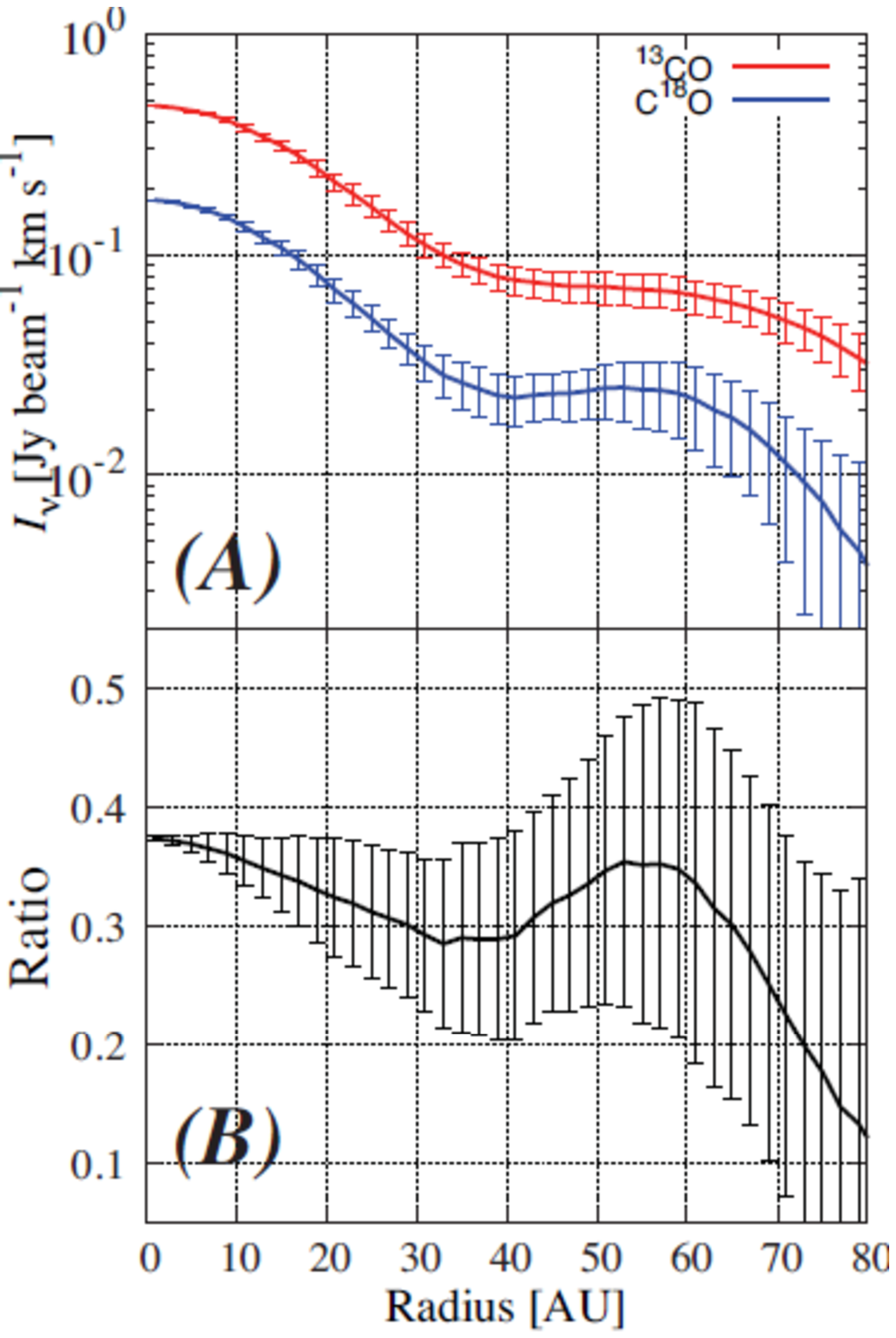}{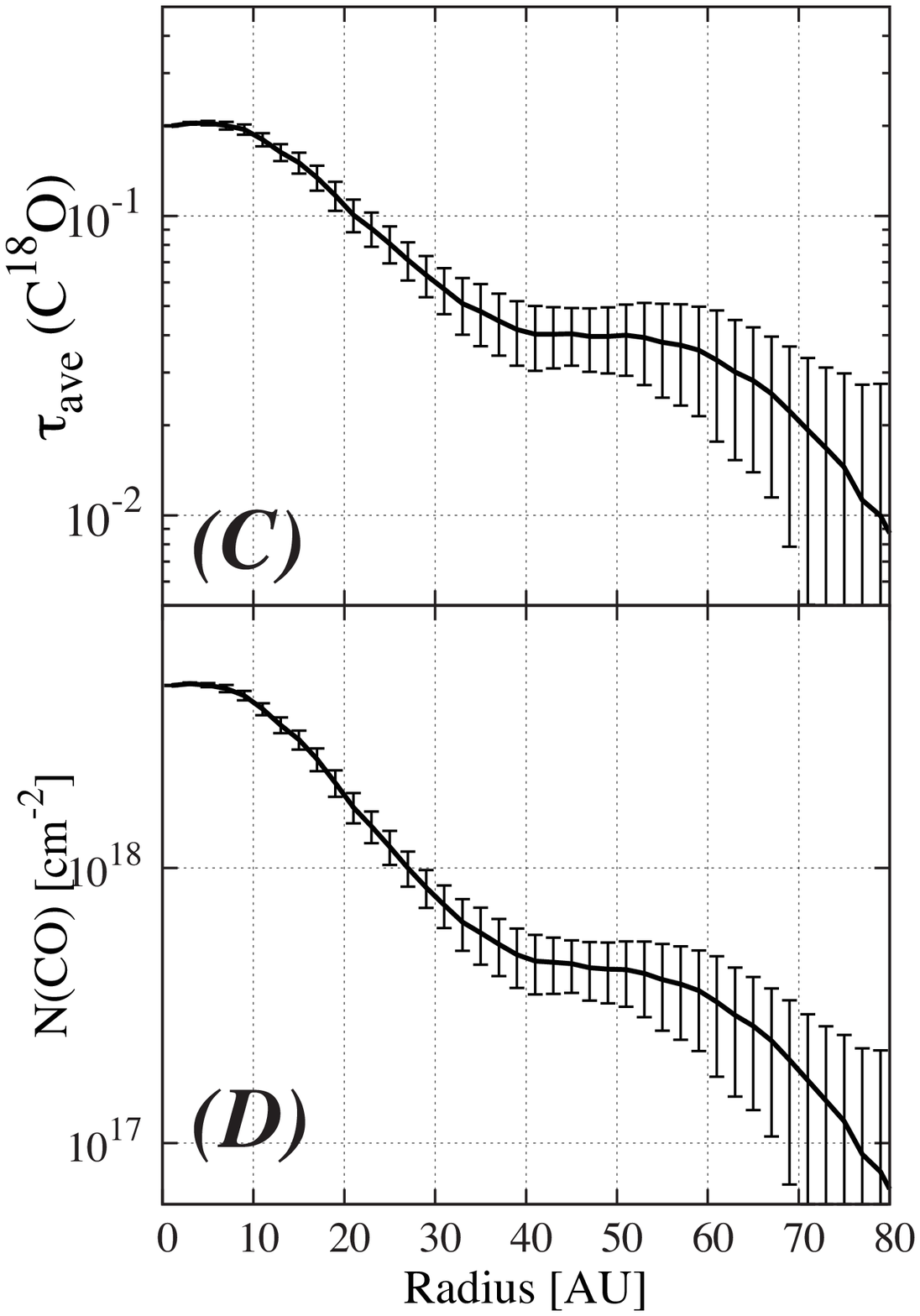}
\includegraphics[scale=.3]{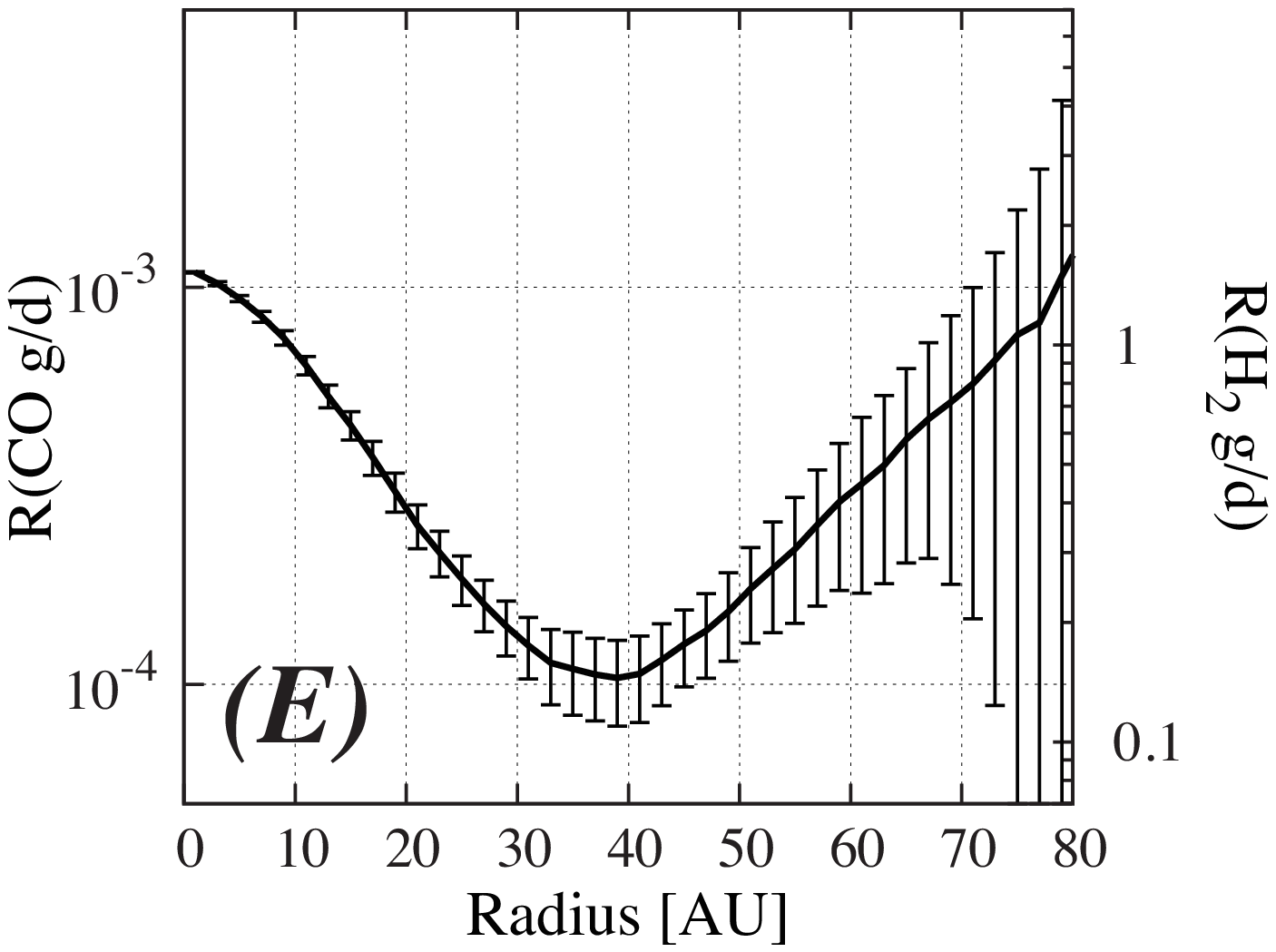}
\caption{Radial distributions of (A) integrated line emission of
 $^{13}$CO ({\it red line}) and C$^{18}$O ({\it blue line}) $J=3-2$, (B)
 line ratio of C$^{18}$O to $^{13}$CO,
 (C) averaged optical depth of the C$^{18}$O line, 
 (D)  CO column density and
 (E) CO gas-to-dust surface density ratio. H$_2$ gas-to-dust ratio is
 also marked for a reference by simply assuming the abundance 
 ratio of CO to H$_2$ of $6\times 10^{-5}$. All data are smoothed to
 a beam size of 0.''45$\times$0.''45. 
}
\end{figure}

\subsection{CO Gas-to-Dust Surface Density Ratio}

Using the obtained CO column density and dust surface density (Figs.~4d
and 2b),
we derive the CO gas-to-dust surface density ratio (Fig.~4e). 
If we convert it
to the H$_2$ surface density, assuming an abundance 
ratio of CO to H$_2$ of $6\times 10^{-5}$ \citep{Qi11}, 
the estimated H$_2$ gas mass is orders of magnitude lower than that
predicted from the 
observations of the HD line emission by the Herschel Space Observatory
\citep{Bergin13}. 
The resulting H$_2$ gas-to-dust surface density ratio ($\sim 0.1-1$) is
about two orders 
of magnitude lower than the typical interstellar value of $\sim$100,
which suggests strong CO depletion throughout the disk down to $\sim 10$AU. 

The CO gas-to-dust surface density ratio increases beyond 
a radius of $\sim 40$AU since the dust surface density drops 
dramatically in this region. It could be due
to drift of pebbles from the outer disk toward the central star 
\citep[e.g,][]{Takeuchi05, Andrews12, Walsh14b}.

\section{Discussions}

\subsection{Origin of the Gap and Ring in the Dust Continuum Emission}

The gap and ring resemble those in the HL Tau system, recently found by 
the ALMA long baseline campaign \citep{ALMA15}. Our result shows that gaps 
and rings in the (sub)millimeter dust continuum can exist, not only in 
relatively young disks ($0.1-1$Myrs) but also 
in relatively old disks ($3-10$Myrs).
One possible mechanism to open a gap is the gravitational interaction 
between a planet and the gas 
\citep[e.g.,][]{Lin79, Goldreich80, Fung14}. 
{Such an interaction may also produce
the spiral density waves recently found in optical and 
near-infrared scattered light imaging of dust grains in protoplanetary 
disks} \citep[e.g.,][]{Muto12}. 
According to recent theoretical analyses of gap structure around a planet
\citep{Kanagawa15a, Kanagawa15b, Kanagawa16}, the depth and width of 
the gap are controlled by the 
planetary mass, the turbulent viscosity and the gas temperature.
The shape of the gap is strongly influenced by 
angular momentum transfer via turbulent viscosity and/or
instability caused by a steep
pressure gradient at the edges of a gap.
The observed gap has an apparent width and depth of 
$\Delta_{\rm gap}\sim 15$AU
and $(\Sigma_0-\Sigma)/\Sigma_0\sim 0.23$, respectively.
This is too shallow and too wide compared with that predicted by 
theory. However, 
the observations are limited to an angular resolution of $\sim 15$AU, 
and the depth and width could be deeper and narrower in reality. 
For instance, if we assume that the gap
depth times the gap width retains the value derived from the observations,
it is possible for the gap to have a width and depth of 
$\Delta_{\rm gap}\sim$6AU and $(\Sigma_0-\Sigma)/\Sigma_0\sim 0.58$,
which is similar to the GPI result \citep{Rapson15}.
Such a gap could be opened by a super-Neptune-mass 
planet, depending on parameters of the disk, such as the turbulent 
viscosity \citep{Kanagawa15a, Kanagawa15b, Kanagawa16}. 
If the gap in the larger dust grains is deeper than that in the gas, 
the planet could be lighter than super-Neptune mass.  
We note that a planet of even a few Earth masses, although it cannot open a gap in
the gas, can open a gap in the dust distribution if a certain amount of
pebble-sized particles, 
whose motion is not perfectly
coupled to that of gas, are
scattered by the planet and/or the spiral density waves excited by the planet 
\citep{Paardekooper06, Muto09}. 
%
%

Another possible mechanism to form a gap and an associated ring in dust continuum
emission is the microscopic process of sintering of CO ice on dust
aggregates \citep{Okuzumi15, Shirono11}. 
The gaps and rings observed in the younger and more luminous HL Tau
system could be explained by 
the sintering of various molecular ices in the disk 
at their distinct snow line locations \citep{Okuzumi15}.
Although our observations indicate that the CO depleted region is 
located down to $\sim 10$AU in the TW Hya system,
model calculations of the temperature in the disk suggest the CO
snow line is located at $\sim 30$AU. 
Sintering is a process that renders an aggregate less sticky
\citep{Shirono11}. Just outside the CO snow line where 
sintering occurs,
large aggregates become easily destroyed into small fragments by collisions 
and their radial drift motion by the above-mentioned mechanism slows down.
Thus dust grains are stuck just outside the CO snow line, and a bright 
ring and a dark lane inside the ring is formed
in the dust continuum emission. 
If another sintering region is formed inside the dark lane by, for
example, CH$_4$ sintering, the region between the bright CO and CH$_4$
sintering regions would look like a gap. 
According to model calculations \citep{Okuzumi15},
the CH$_4$ sintering region is located
at $\sim 10$AU in the TW Hya disk. 

%
%

\subsection{CO Gas Depletion inside the CO Snow Line}

From the CO line observations, we find a very low column density of CO
compared with dust throughout the disk down to a radius of about 10AU. 
{This CO depletion could indicate a general absence of H$_2$ gas
compared with dust. However, the Herschel HD observations
\citep{Bergin13} indicates that it is more likely that 
it is due to CO freeze-out on grains with the possibility that
subsequent grain surface reactions form larger molecules 
even inside the CO snow line \citep{Favre13, Williams14}}. 
The CO column density derived from our observations, $N_{\rm CO}\sim
10^{18}$cm$^{-2}$, 
can be
explained by detailed model calculations using a chemical
network which includes freeze-out of molecules on grains and 
grain-surface reactions \citep{Aikawa15}. 
{The model calculations show the CO depletion will proceed inside the CO
snow line due to the sink effect: conversion of CO to less volatile
species on grain surface. In the case of TW Hya, it could occur
on a timescale shorter than the disk lifetime, depending on
the amount of small grains. See \cite{Aikawa15}
for more detailed discussions including the effect on other species.}
{The CO depletion spread over the disk is inconsistent with the prediction 
by the previous ALMA N$_2$H$^+$ observations that 
depletion would be localized beyond the CO snow line} \citep{Qi13}. 
This could be because the N$_2$H$^+$ line
emission traces the disk surface and not the CO depleted region.

\begin{figure}
\plotone{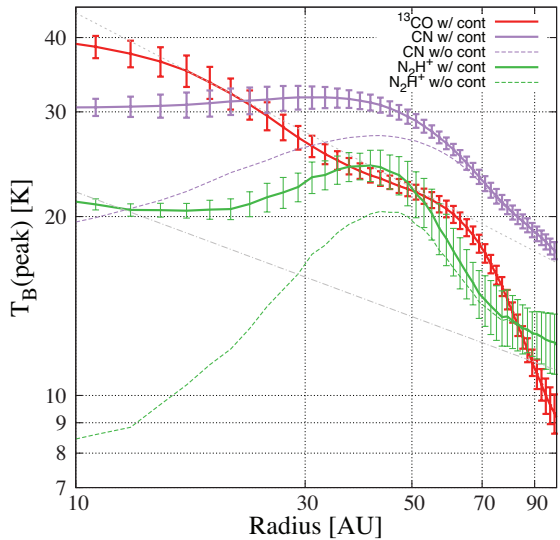}
\caption{Radial distributions of the observed brightness temperature at
 the peak of $^{13}$CO $J=3-2$ (red) and N$_2$H$^+$ $J=4-3$ (green)
 lines. The dust temperature at the midplane (dot-dashed gray line) 
 and the brightness temperature of CN $N=3-2$ (purple) line are also plotted 
 for comparison. 
Dashed lines show the data
 subtracted by the dust continuum emission for N$_2$H$^+$ and CN.
All data  are smoothed to a beam size of 0.''45$\times$0.''45.
}
\end{figure}

Fig.~5 shows the brightness temperature at the peak of the $^{13}$CO
$J=3-2$ line obtained by our observations and the N$_2$H$^+$ $J=4-3$
line at 372.672GHz obtained by the ALMA archived data (2011.0.00340.S).
%
Since the $^{13}$CO line is optically thick, 
the brightness temperature at the peak of the line emission
represents the gas temperature of the line emitting region. 
%
If LTE is applicable, the N$_2$H$^+$ brightness temperature is 
higher than the gas temperature of the $^{13}$CO line emitting region at
the disk radius of $\sim 40$AU, and higher than the dust temperature
near the midplane down to $\sim 15$AU (Fig.~5). If the N$_2$H$^+$ line is 
optically
thin, the gas temperature of the line emitting region is higher than the
brightness temperature. Therefore, the N$_2$H$^+$ line should come from
the surface layer of the disk. 
%
Model calculations also predict that N$_2$H$^+$ exists in the disk
surface for the model with (sub)micron-sized grains,
similar to {radical species abundant in the disk surface, such as CN and
C$_2$H \citep[e.g.,][]{Walsh10, Aikawa15}. Our results (Fig.~5) and
the SMA observations of C$_2$H \citep{Kastner15} show that the radial
intensity profiles of these species are similar. They have peaks around
the disk radius of 40AU beyond which the dust surface density drops.} 
The result suggests that in order to trace the CO depleted
region, the C$^{18}$O line may be more robust than the N$_2$H$^+$ line.

In the CO depleted region complex organic molecules would
be produced via grain surface reactions since hydrogen attachment to
CO is thought to produce methanol and more complex species 
\citep[e.g.,][]{Watanabe08, Walsh14a}.
Methyl cyanide, which has recently been detected from a protoplanetary disk
for the first time by ALMA \citep{Oberg15},
could be formed through such grain surface reactions.  

\acknowledgments

We would like to thank the referee for his/her comments which improved
our paper. We are also grateful to Sean Andrews and Joel Kastner for
their fruitful comments.
This paper makes use of the following ALMA data: ADS/JAO.ALMA\#2013.1.01397.S
and ADS/JAO.ALMA\#2011.0.00340.S.
ALMA is a partnership of ESO (representing its member states), NSF (USA) and NINS
(Japan), together with NRC (Canada), NSC and ASIAA (Taiwan) and KASI (Republic of Korea), in cooperation with
the Republic of Chile. The Joint ALMA Observatory is operated by ESO, AUI/NRAO
and NAOJ. This work is partially supported by Grants-in-Aid for Scientific Research, 23103005, 25108004, 25400229 and 15H03646. 
T.T. was supported by JSPS KAKENHI grant No. 23103004.
C.W. is supported by the Netherlands Organisation for Scientific
Research (program number 639.041.335).
Astrophysics at QUB is supported by a grant from the STFC.



{\it Facility:} \facility{ALMA}.

\clearpage

\end{document}